\documentstyle[preprint,aps]{revtex}

\begin{document}

\title
{Low-energy spectra in the $t$-$J$ type models at light doping} 
\author{Wei-Cheng Lee$^{1}$, T.~K. Lee$^{1,2}$, Chang-Ming Ho$^{2}$ and
P.~W. Leung$^{3}$}
\address{$^1$ Institute of Physics, Academia Sinica,
Nankang, Taipei, Taiwan 11529\\
$^2$ Physics Division, National Center for Theoretical Sciences,
P.O.Box 2-131,Hsinchu, Taiwan 300\\
$^3$ Physics Departmewnt, Hong Kong University of
Science and Technology, Clear Water Bay, Hong Kong}

\date{\today}

\maketitle

\begin{abstract}
Based on a variational approach, we propose that there are two
kinds of low energy states in the
$t$-$J$ type models at low doping.
In a quasi-particle state
 an unpaired spin  bound to a hole with a well-defined momentum
can be excited with spin waves. The resulting state shows a suppression of
antiferromagnetic order around the hole with the profile of a {\em spin bag}. 
These spin-bag states with spin and charge or hole separated 
form a continuum of low-energy excitations. 
Very different properties predicted by these two kinds of states explain 
a number of anomalous results observed in the exact diagonalization 
studies on small clusters up to 32 sites. 
\end{abstract}

\pacs{PACS numbers: 74.72.Jt, 75.50.Ee, 79.60.-i}

Soon after the discovery of the high $T_c$ superconducting 
cuprates, 
 the  $t$-$J$ model was proposed\cite{pwa87}
as the prototype to examine the 
phenomena. 
Since then it has been shown in theories\cite{many,lhn} and 
experiments\cite{Na-CCOC,dam} 
that extra 
hoppings beyond nearest neighbors ($n.n.$)
are also important 
to the low-energy 
features of the cuprates.
Enormous theoretical effort has been devoted to predict 
 the low-energy spectra of these $t$-$J$ type models. 
But the strong correlation  
associated with these models has made
perturbative approaches ineffective. 
With the many different mean-field theories presented so far, 
there is 
little consensus to the "correct" description of low-energy spectra 
of these models. To sort out the proper theory it is important to first 
have a detail comparison of the predictions with the results of exact 
calculations.

Recently, significant progress has been made at very low doping. Several 
different numerical techniques, such as 
exact diagonalization (ED) studies\cite{toh-mae94,zk95,zk97,dagotto},
self-consistent Born approximation (SCBA)\cite{mar} and
the Green function Monte Carlo method\cite{dago2}, all have obtained
similar results for the energy-momentum dispersion relation of a single
hole doped into $t$-$J$ type models. The results
agree fairly well 
 with experiments\cite{dam}. In addition, 
a  mean-field or variational wave function (VWF)\cite{lee-shih}
constructed from the half-filled
Mott insulating state with antiferromagnetic long-range order (AF LRO) 
has also 
obtained a similar success. In a recent paper\cite{lhn} this single-hole VWF 
is generalized to treat systems with multiple holes
or electrons. The new set of VWF's easily explained 
 angle-resolved photoemission spectroscopy (ARPES)
results\cite{Na-CCOC} for Fermi pockets
around ($\pi/2$,$\pi/2$) and ($\pi$,0) for hole-doped
and electron-doped systems, respectively.
 It\cite{lhn} also reproduced unusual patterns in momentum distribution
functions (MDF's) of the ground state
 calculated by the ED method\cite{2-in-32,leung}
for one and two holes in 32 sites.
For these VWF's, doped holes or electrons behave like quasi-particles (QP's). 
In this paper  we will show that in addition to these 
QP states, there is a continuum of charge excitations 
described by {\em spin-charge separated} states in the spectra
of $t$-$J$ type models. The presence of the two kinds of 
states is carefully examined by explaining
several anomalous results reported by ED studies for clusters up to 32 sites. 

As shown by Lee {\em et al.}\cite{lhn}, the ground state in the presence of a 
few doped holes or electrons could be described by a VWF constructed from the 
half-filled Mott insulating state. At half-filling, the ground state of the
Heisenberg Hamiltonian
is described fairly accurately by including three
mean-field paramters\cite{lee-shih,lee-feng}: 
the staggered magnetization
$m_s$=$\langle S^z_A\rangle$=$-\langle S^z_B\rangle$, where the lattice
is divided into {\em A} and {\em B} sublattices; the uniform bond order 
parameters 
$\chi$=$\langle \sum_{\sigma}c^{\dagger}_{i\sigma}c_{j\sigma}\rangle$; 
and $d$-wave resonating valence bond ($d$-RVB) order
$\Delta$=$\langle c_{j\downarrow}c_{i\uparrow}-c_{j\uparrow}c_{i\downarrow}
\rangle$ if $i$ and $j$ are $n.n.$ sites in $x$-direction and $-\Delta$
for $y$-direction. Without $d$-RVB order, the mean-field Hamiltonian has 
lower and upper spin-density-wave (SDW) bands with operators 
$a_{{\bf k}\sigma}$=$\alpha_{\bf k}c_{{\bf
k}\sigma}+\sigma\beta_{\bf k}c_{{\bf k}+{\bf Q}\sigma}$, and
$b_{{\bf k}\sigma}$=$-\sigma\beta_{\bf k}c_{{\bf
k}\sigma}+\alpha_{\bf k}c_{{\bf k}+{\bf Q}\sigma}$, respectively. Here
${\bf Q}$=$(\pi,\pi)$, 
$\alpha_{\bf k}^2$=${1\over 2}$[$1-(\epsilon_{\bf k}/\xi_{\bf k}$)] and 
$\beta_{\bf k}^2$=${1\over 2}$[1+($\epsilon_{\bf k}$/$\xi_{\bf k}$)]. 
Energy dispersions for the two SDW bands are
$\pm\xi_{\bf k}$=
$\pm(\epsilon_{\bf k}^2+(Jm_s)^2)^{1\over2}$ with
$\epsilon_{\bf k}$=$-{3 \over 4}J\chi (cos{\rm k}_x+cos{\rm k}_y)$.
Inclusion of $d$-RVB pairing for electrons on total $N_{s}$ sites, the 
VWF for the ground state has the form 
$|\Psi_0\rangle$=$P_e [{\sum_{\bf k} (A_{\bf k}
a^{\dagger}_{{\bf k}\uparrow}a^{\dagger}_{{\bf -k}\downarrow}+B_{\bf k}
b^{\dagger}_{{\bf k}\uparrow}b^{\dagger}_{{\bf -k}\downarrow})}]^{N_s/2}
|0\rangle $,
where $A_{\bf k}$=$(E_{\bf k}+\xi_{\bf k})/\Delta_{\bf k}$ and
$B_{\bf k}$=$-(E_{\bf k}-\xi_{\bf k})/\Delta_{\bf k}$ with
$E_{\bf k}$=($\xi_{\bf k}^2+\Delta_{\bf k}^2$)$^{1/2}$ and the 
constraint of one electron per site enfoced by $P_e$. Here $\Delta_{\bf k}$
=${3\over 4}J\Delta d_{\bf k}$ with $d_{\bf k}$=$cos{\rm k}_x-cos{\rm k}_y$.
Notice that the sum in $|\Psi_0\rangle$ is taken over the sublattice
Brillouin zone (SBZ). 

In the presence of doped holes or electrons, we consider the $t$-$J$ type model
including longer-ranged hoppings, with amplitues $t'$ for the 2nd {\em n.n.} 
and $t''$ the 3rd {\em n.n.}. 
By applying a particle-hole 
transformation\cite{lhn,toh-mae94} we can treat hole- and 
electron-doped cases in the same manner. However, here we will 
just concentrate on the hole-doped cases with  $J/t$=$0.3=-t'/t$ 
and $t''/t$=$0.2$.

When a hole is doped or an electron is removed from $|\Psi_0\rangle$,
a pair must be broken with an unpaired spin left. Thus it is quite
natural to have the following VWF\cite{lhn,lee-shih} for a single doped hole, {\em e.g.}, 
with a lone up spin
\begin{eqnarray}
|\Psi_1({\bf q}_{s})\rangle & = &
P_d~c^{\dagger}_{{\bf q}_{s}\uparrow} \nonumber \\
& & \mbox{} [{\sum_{[{\bf k} \neq {\bf q}_{h}]} 
(A_{\bf k} a^{\dagger}_{{\bf k}\uparrow}a^{\dagger}_{{\bf -k}\downarrow}
 +B_{\bf k} b^{\dagger}_{{\bf k}\uparrow}b^{\dagger}_{{\bf -k}\downarrow})}
]^{(N_s/2)-1}
|0\rangle ,  \nonumber
\label{twf-sb}
\end{eqnarray}
where the hole momentum ${\bf q}_{h}$ is excluded from the sum
if ${\bf q}_{h}$ is within the  SBZ,
otherwise, ${\bf q}_{h}-{\bf Q}$ is excluded. 
$P_d$ here enforces the constraint of no doubly occupied sites. 
When we choose 
the unpaired-spin momentum ${\bf q}_{s}$ to be either
the same as the hole momentum ${\bf q}_{h}$ or ${\bf q}_{h}$+${\bf Q}$,
this VWF is equivalent to the Lee-Shih\cite{lee-shih} wave function.
Variational energies calculated vary with  ${\bf q}_{h}$\cite{lhn,lee-shih}. 
The energy dispersions for $t$-$J$ and $t$-$t'$-$t''$-$J$ models are plotted 
as filled circles in Fig. 1(a) and (b), respectively. For both models, 
the ground state with one hole
has momentum $(\pi/2,\pi/2)\equiv {\bf Q}/2$. As shown in Ref.\cite{lhn},
these dispersion relations are still followed when hole number is increased.  
The holes in these wave functions behave just like QP's, hence
we denote $|\Psi_1({\bf q}_{h}$=${\bf q}_{s})\rangle$$\equiv$$|\Psi_1^{qp}\rangle$.

There are only two variational
parameters: $\Delta/\chi$ and $m_s/\chi$ in our VWF's.
The extended hoppings  $t'$ and $t''$ are not used as variational 
parameters in
both $|\Psi_1^{qp}\rangle$ and $|\Psi_1\rangle$.
The effect of $t$ is included in the RVB uniform bond 
$\chi$. Since  $t'$ and $t''$ are compatible with AF LRO, 
there is little effect for them to be included in our VWF.  

Clearly, the choice of  unpaired spin to have the same momentum
as the hole, {\em i.e.} ${\bf q}_{s}$=${\bf q}_{h}$, is a special case for 
$|\Psi_1\rangle$. If we choose ${\bf q}_{s}\neq{\bf q}_{h}$, then 
not only the electron pair at ${\bf q}_{h}$ and $-{\bf q}_{h}$ is 
excluded in the sum in $|\Psi_1\rangle$, the pair at  ${\bf q}_{s}$ 
and $-{\bf q}_{s}$ is also affected. Hence we expect it to be higher in energy.
To make a distinction from the afore-mentioned QP
states  $|\Psi_1^{qp}\rangle$, these states will be denoted as 
the {\em spin-bag} (SB) states, {\em i.e.} $|\Psi_1({\bf q}_{s}\neq{\bf q}_{h})\rangle \equiv |\Psi_1^{sb}\rangle$. 
The variational energies as a function of ${\bf q}_{s}$ 
for $|\Psi_1^{sb}\rangle$ with ${\bf q}_{h}$=${\bf Q}/2$ are plotted as 
empty circles in Figs. 1(a) and (b) for the $t$-$J$ 
and $t$-$t'$-$t''$-$J$ models, respectively. Many SB states could be 
constructed with the same ${\bf q}_{s}$ but different ${\bf q}_{h}$. 
While it is possible to have the SB states of even lower energy with 
${\bf q}_{h}$=$(3{\pi}/4,0)$ in the $t$-$J$ model (gray circles in Fig. 1(a)), 
they are of higher energies than that of 
SB states with ${\bf q}_{h}$=$(\pi/2,\pm\pi/2)$ in the 
$t$-$t'$-$t''$-$J$ case. There are many such states, actually an infinite 
number of them for an infinite system, forming a continuum, as schematically 
illustrated by  the shaded regions in Fig. 1(a) and (b). 

There is an intuitive way to understand the difference between SB and QP states.
The spin excitations
of the QP states can be easily constructed 
by applying the spin operators,   
$S^{\dagger (-)}({\bf k}')$=$\sum_{{\bf q}'} c^{\dagger}_{{\bf q}' + {\bf k}'\uparrow(\downarrow)} c_{{\bf q}' \downarrow(\uparrow)}$,  
to $|\Psi_1^{qp}\rangle$. Notice that these operators
commute with the projection operator $P_d$. In the linear spin-wave theory 
${\bf k}'$ is the momentum of the spin wave.
 The particular term inlcuded in the sum
of ${\bf q}'$ with ${\bf q}'$ equal to the 
momentum of the unpaired spin  ${\bf q}_{s}$=${\bf q}_{h}$ changes the
QP state $|\Psi_1^{qp}\rangle$ 
to the SB ones $|\Psi_1^{sb} ({\bf q}_{s}$=${\bf q}_{h} + {\bf k}')\rangle$. 
Thus the SB states are actually just spin-wave excitations of
the QP state with the same hole mementum.
In Fig. 1(c) the  difference of variational energies between the SB states
and QP state with ${\bf q}_{h}$=${\bf Q}/2$
for the $t$-$t'$-$t''$-$J$ model is plotted
as a function of the difference of momentum 
${\bf k}'$=${\bf q}_{s}-{\bf q}_{h}$. The dotted line is the prediction of
energy-momentum dispersion relation of linear spin-wave theory\cite{sw}. 
The slight differences between the two results at ${\bf k}'$=${\bf Q}/2$ 
and ${\bf k}'$=$(\pi,0)$ are due to the hole-renormalization effect\cite{ho}. 
SB states represent spin excitations of the QP states.
The empty/gray circles in Figs. 1(a) and (b) are the lowest spin excitation 
energies of the ground state\cite{nt}.

The two VWF's also give very different spin and hole correlations. When the hole is at the 
$A$ sublattice, the correlation is 
$SH_{A}({\bf r})$=$\sum_{i \in A}(-1)^{i+{\bf r}}
\langle \Psi_{1} | n_i^h S^{z}_{i+{\bf r}}  |\Psi_{1} \rangle / \sum_{i \in A}\langle \Psi_{1}| n_i^h|\Psi_{1} \rangle$, 
where $n_i^h$ is the hole-number operator at site $i$; similarly for the case when hole is at
$B$ sublattice.  In our convention, the up spin prefers to be on
sublattice $A$. If the hole and unpaired spin are uncorrelated, $SH_{A,B}$
equals to the value of uniform staggered magnetization. 
Values of $SH_A$ for
the QP state (filled triangles) with ${\bf q}_{s}$=${\bf q}_{h}$=$(0,0)$ and
the SB state (empty triangles) with ${\bf q}_{s}$=$(0,0)$ and  
${\bf q}_{h}$=${\bf Q}/2$\cite{note1}, respectively, are plotted as a 
function of distance from the hole in Fig. 2. Both states have an 
unpaired down spin and a single hole. 
The spin configurations around the hole are clearly different 
for the two states. The spin magnetization right next to the hole in the 
QP state $|\Psi_1^{qp} \rangle$ has values larger than the uniform 
background, $0.368$. Thus the unpaired spin is bound to the hole. On 
the other hand, for the SB state the magnetization is suppressed around 
the hole, this is similar to the idea of a SB first proposed in Ref.\cite{sch}. The 
unpaired spin bound to the hole in the QP state is here being excited by the 
spin-wave excitation and becomes unbound in the SB state. This may be viwed 
as a spin-charge separated state.

The spin-charge separation observed in the SB states has 
many interesting consequences.
In ED studies, it has been found\cite{eder} that the lowest energy state
at $(\pi,0)$ for the $t$-$t'$-$t''$-$J$ model is very 
different from that of the $t$-$J$ model. The spin-spin 
correlation across the hole changes from
ferromagnetic (FM) to AF when $t'$ is turned on. This result can now 
be understood as that the lowest 
energy state at $(\pi,0)$ for the $t$-$J$ model is
the QP state $|\Psi_1^{qp}({\bf q}_{s}$=${\bf q}_{h}$=$(\pi,0))\rangle$ 
as shown in Fig. 1(a), but it 
changes to a 
SB state $|\Psi_1^{sb}({\bf q}_{s}$=$(\pi,0),{\bf q}_{h}$=${\bf Q}/2)\rangle$ for the $t$-$t'$-$t''$-$J$ 
model as shown in Fig. 1(b). In Table I we list correlations obtained between pairs of spins around 
the doped hole for QP and SB states. The correlation is defined as
$C_{\delta,\delta'}({\bf q}_{s})
\equiv \sum_i \langle \Psi_{1}^{qp,sb}({\bf q}_{s})| n_i^h {\bf S}_{i+\delta} \cdot {\bf S}_{i+\delta'} |
\Psi_{1}^{qp,sb}({\bf q}_{s}) \rangle$ with $\delta$ and $\delta'$ denoting 
two sites around the hole\cite{eder}.
While the lower-energy QP states at ${\bf q}_{s}$=${\bf q}_{h}$=($\pi,0$) 
and (0,0) behave as expected for a system with AF LRO, {\em i.e.} 
FM for spins at the same sublattice (pairs {\bf a} and {\bf b}) and AF otherwise, the result for a SB state 
with ${\bf q}_s$=$(\pi,0)$ and ${\bf q}_{h}$=${\bf Q}/2$ shows AF correlation at the same sublattice 
This is exactly the behavior observed in ED results\cite{eder}. Our result in Fig. 1 shows that at (0,0) the lowest energy state
remains to be the QP state even when $t'$ and $t''$ are included.
The spin correlations are thus not changed, this is also consistent with what is found by Tohyama {\em et al.}\cite{eder}.   
 All these results clearly support our identification of 
low-energy spectra to be consisted of QP and SB states. 


\vspace{0.5cm}

{\bf TABLE I.} $C_{\delta,\delta'}({\bf q}_{s})$ calculated for an 
8$\times$8 lattice using different VWF's with momenta indicated. The first and third rows are for QP states while others are for SB ones.
{\bf a} to {\bf e} are different pairs of sites defined in the 
inset of Fig. 2. 
Positive(negative) values mean 
FM(AF) correlations. 

\begin{center} 
\begin{tabular}{c | c | c c c c c} \hline
\hline
${\bf q}_{s}$ & ${\bf q}_{h}$ & {\bf a} & {\bf b} & {\bf c} & {\bf d} & {\bf e}\\
\hline
(0,0) & (0,0) & 0.188 & 0.188 & 0.202 & -0.273 & -0.264 \\
\hline
(0,0) & {\bf Q}/2 & -0.0288 & -0.0254 & -0.0302 & -0.203 & -0.195 \\
\hline
($\pi,0$) & ($\pi,0$) & 0.123 & 0.15 & 0.071 & -0.353 & -0.279 \\
\hline
($\pi,0$) & {\bf Q}/2 & -0.0313 & -0.0085 & -0.002 & -0.1921 & -0.212\\
\hline
\end{tabular}
\end{center}
\vspace{0.5cm}


Another important difference between QP and SB states is in their
MDF $\langle n_{\sigma}({\bf k})\rangle$. The results can actually be predicted. Since the hole momentum
 ${\bf q}_{h}$ is excluded from the VWF  $|\Psi_1\rangle$, we naturally expect
 $\langle n_{\uparrow}({\bf k})\rangle$ to have a smaller
value or a dip at ${\bf k}$=${\bf q}_{h}$ and ${\bf k}$=${\bf q}_{h}+{\bf Q}$
than its neighbors, similarly for $\langle n_{\downarrow}({\bf k})\rangle$
at ${\bf k}$=$-{\bf q}_{h}$ and  ${\bf k}$=$-{\bf q}_{h}+{\bf Q}$. 
However, in a QP state with an up spin at momentum ${\bf q}_{s}$=${\bf q}_{h}$, 
then $\langle n_{\uparrow}({\bf k}$=${\bf q}_{h})\rangle$ is increased 
and there is no more  a dip. As an example,  the MDF obtained by 
the QP state with  ${\bf q}_{h}$=${\bf q}_{s}$=$(\pi,0)$ is listed in 
Fig. 3(a). Because of the symmetry, only results for one quadrant of 
the  BZ are shown. At each ${\bf k}$ the upper(lower) number is for up(down) spin.
Now for a SB state with a lone up spin at ${\bf q}_{s}(\neq{\bf q}_{h})$, 
the original spin at ${\bf q}_{h}$ in the QP state is excited and placed 
at  ${\bf q}_{s}$, then $\langle n_{\uparrow}({\bf k}$=${\bf q}_{h})\rangle$ 
still has a dip. Fig. 3(b) shows the MDF obtained for the SB 
state with ${\bf q}_{s}$=$(\pi,0)$ and ${\bf q}_{h}$=${\bf Q}/2$\cite{note1}.
Results are here for 32 sites and $(\Delta/\chi,m_{s}/\chi)$=$(0.1,0.05)$.

This behavior of the MDF's is indeed found in the exact results of 
the $t$-$t'$-$t''$-$J$ and $t$-$J$ models on 32 sites. MDF's obtained 
by the ED method for the lowest energy state at $(\pi,0)$ of the $t$-$J$  
and $t$-$t'$-$t''$-$J$ models are shown in Fig. 3(c) and (d), respectively.
The nice qualitative agreement achieved between 3(a) and (c) as well as 
between 3(b) and (d)\cite{note2} re-affirms our results shown 
in Fig. 1(a) and (b): the lowest energy state at $(\pi,0)$ for 
a single hole is a QP state for the $t$-$J$ model and a SB state 
for the $t$-$t'$-$t''$-$J$ model. 

Another consequence of this switch from a QP state to a SB one is the 
drastic change of the spectral weight,
$Z_{\bf k}$=$|\langle \Psi_{1}({\bf k}) | c_{{\bf k}\sigma} | \Psi_{0}
\rangle|^{2}/ \langle \Psi_{0} | c^{\dagger}_{{\bf k}\sigma}
c_{{\bf k}\sigma} | \Psi_{0} \rangle$. 
Using $|\Psi_1^{qp}({\bf q}$=$(\pi,0))\rangle$ 
we obtained $Z_{\bf k}=0.475$, it vanishes when we use 
the SB state $|\Psi_1^{sb}({\bf q}_{s}$=$(\pi,0),{\bf q}_{h}$=${\bf Q}/2)\rangle$. 
This is consistent with exact results for the $t$-$J$ model ($Z_{\bf
k}$=$0.34$)\cite{zk95} and $t$-$t'$-$t''$-$J$ model ($Z_{\bf k}$=0)\cite{footnote}. 
In addition, spectral weights of the lowest energy states of both models
at $(\pi,\pi)$ and $(3\pi/4,3\pi/4)$ are either exactly zero or very
small. This is consistent with our identification that states at both momenta are SB ones. 
Since $c_{{\bf k}\sigma} | \Psi_{0}\rangle$, unlike the SB state, has momenta of the 
hole and  unpaired spin related, it has a negligible 
overlap with the SB state. By contrast, states at $(\pi/2,\pi/2)$ 
and $(\pi/4,\pi/4)$ remain to be QP states in both $t$-$J$ 
and $t$-$t'$-$t''$-$J$ models, hence large spectral weights are expected. 
It is noted
that our QP (SB) states predict much larger (small) spectral weights 
in comparison with that of the exact 32 sites. This discrepency is partly due to the fact 
that we have AF LRO in our VWF's while total spin is a good quantum 
number in the exact results. Another reason is that due to the projection 
operator $P_d$ 
our QP states and SB states with the same quantum numbers
(total momentum and total $S_z$) are actually not orthogonal to each other although 
they have very small overlap.
But there are many SB states in the continuum that could couple with 
a particular QP state. Hence, when the QP state has energies very close to the 
continuum, the spectral weight of the QP state is diluted by the
coupling with SB states. This effect makes the quantitative prediction of  
spectral weight  difficult. We leave this issue for the future work.

In summary, based on a mean-field theory with AF and $d$-RVB order parameters 
we have proposed that at low doping
there are two kinds of low energy states for $t$-$J$ type models. The 
single-hole QP states have a well-defined energy dispersion. 
By exciting the QP states with spin waves we obtain a continuum of
SB states. The unpaired spin is separated from the hole in the SB states.
A number of physical properties predicted by these two kinds of states
are in good agreement with the exact results obtained by ED studies.
Although our emphasis in this paper is to show the solid theoretical 
support of these two kinds of states, 
there are also experimental evidences. 
In Ref.\cite{lhn}, 
QP states were shown to explain well the single-hole 
dispersion observed by ARPES. However the 
overall variation pattern of spectral weights
observed in the ARPES experiment on $Ca_{2}CuO_{2}Cl_{2}$\cite{Na-CCOC}
is naturally understood with the presence of SB states: notable lowest 
energy peaks are only observed in small regions of k-space, e.g. near
$(\pi/2,\pi/2)$ and $(\pi/2,0)$ where QP states have 
lower energy than the SB states (Fig.1(b)). More
comparison with experiments is in progress.

     

\noindent

We are grateful to Profs. N. Nagaosa and C.-T. Shih for invaluable discussions and their 
supporting data. TKL is supported by the grant NSC91-2112-M-001-011 (R.O.C.). 
PWL 
is supported by the Hong Kong RGC grant number HKUST6159/01P.


\newpage


\begin{figure}
\caption{Variational energies calculated for the (a) $t$-$J$
 and (b) $t$-$t'$-$t''$-$J$ model Hamiltonians for one hole 
on an 8$\times$8 lattice by applying our TWF's. Filled circles, 
connected by solid lines, are VMC results using the 
$| \Psi_{1}^{qp}({\bf q}_{s}) \rangle$ discussed in the text; 
empty(gray) circles by $| \Psi_{1}^{sb}({\bf q}_{s},{\bf q}_{h}$=${\bf Q}/2
$($(3{\pi}/4,0)) \rangle$. 
Shaded regions indicate a possible continuum for an infinite system.
A minus sign has been multiplied to all the data shown here. 
(c) Difference of variational energies between
the QP ground state at ${\bf Q}/2$ and the SB states
at ${\bf q}_{s}$=${\bf Q}/2 + {\bf k}'$
in (b)
as a function of ${\bf k}'$.
The dotted line is the prediction of energy dispersion of 
linear spin-wave theory [17]. Results here are 
obtained with parameters ($\Delta/\chi,m_{s}/\chi$)=(0.25,0.125).}
\label{qp-sb-disp1} 
\end{figure}



\begin{figure}
\caption{Spin-hole correlation functions,  $SH_A$ (defined in the text),
for an 8$\times$8 lattice calculated using the QP state (filled triangles)
with ${\bf q}_{s}$=${\bf q}_{h}$=$(0,0)$ and the SB state (empty triangles)
with ${\bf q}_{s}$=$(0,0)$ and ${\bf q}_{h}$=${\bf Q}/2$ [20].
Inset: pairs of sites, denoted by letters {\bf a} to {\bf e}, where spin-spin
correlations listed in Table I are computed. Parameters are the same as used in Fig. 1.}
\label{sh-qp-sb-00-1}
\end{figure}


\begin{figure}
\caption{Momentum distribution functions in one quadrant
of ${\bf k}$ space 
the  BZ for 32 sites obtained by (a) $|\Psi_1^{qp}({\bf q}$=$(\pi,0))\rangle$, 
(b) $|\Psi_1^{sb}({\bf q}_{s}$=$(\pi,0),{\bf q}_{h}$=${\bf Q}/2)\rangle$, 
(c) ED results for $(\pi,0)$ state of the $t$-$J$ model and (d) $t$-$t'$-$t''$-$J$ model. 
The upper number is result for up spin and lower for down spin.}
\label{nk-32-pi0} 
\end{figure}



\begin{references}
\bibitem{nt}The gap between the continuum and the QP dispersion is an 
artifact of the finite size. For an infinite system, there is no gap.
\bibitem{note2}Due to the assumption of AF LRO, our values of MDF's 
outside of the SBZ are larger than ED results.
\bibitem{footnote} Note that the $(\pi,0)$ state reported in Ref.\cite{zk97} is 
{\it not} the lowest energy state with the same momentum. The lowest one has zero overlap
with $c_{{\bf k}\sigma}|\Psi_0 \rangle$ and, consequently, $Z_{\bf k}$=$0$. 
This state was missing from the spectrum in Ref.\cite{zk97}.
\end{references}
\end{document}